# Modelling and Analysis of Magnetic Fields from Skeletal Muscle for Valuable Physiological Measurements

## Author Information

## Affiliations

**James Watt School of Engineering, University of Glasgow, Glasgow, UK.**

Siming Zuo and Hadi Heidari (z.siming.1@research.gla.ac.uk & hadi.heidari@glasgow.ac.uk)

**School of Informatics, University of Edinburgh, Edinburgh, UK**

Kianoush Nazarpour (kianoush.nazarpour@ed.ac.uk)

**Department of Bioengineering, Imperial College London, London, UK**

Dario Farina (d.farina@imperial.ac.uk)

**Ostschweizer Kinderspital, Neuropediatric Department, Switzerland**

Philip Broser (philip.broser@kispisg.ch)

## Corresponding author

Correspondence to: Hadi Heidari (Tel: 44 (0)141 330 6789), Microelectronics Lab (meLAB) Glasgow

## Abstract

MagnetoMyoGraphy (MMG) is a method of studying muscle function via weak magnetic fields generated from human active organs and tissues. The correspondence between MMG and electromyography means directly derived from the Maxwell-Ampère law. Here, upon briefly describing the principles of voltage distribution inside skeletal muscles due to the electrical stimulation, we provide a protocol to determine the effects of the magnetic field generated from a time-changing action potential propagating in a group of skeletal muscle cells. The position-dependent and the magnetic field behaviour on account of the different currents in muscle fibres are performed in temporal, spectral and spatial domains. The procedure covers identification of the fiber subpopulations inside the fascicles of a given nerve section, characterization of soleus skeletal muscle currents, check of axial intracellular currents, calculation of the generated magnetic field ultimately. We expect this protocol to take approximately 2-3 hours to complete for whole finite-element analysis.

## Introduction

With the rapid development of nanoscale magnetic sensors, the non-invasive biomagnetic recording has been a reliable and robust approach for biomedical applications, ranging from clinical diagnoses to human-computer-interaction[1]. The assessment of muscle activity is essential in nowadays diagnosis of peripheral muscle and nerve diseases, in fundamentals of movement neuroscience and in technologies for motor rehabilitation. Detecting weak magnetic signals derived from human skeletal muscle, first formally proposed in 1972 and it was called Magnetomyography (MMG)[2]. The scientists have recorded the magnetomyogram signal as one component of the magnetic field vector for time at the point of measurement, in which the magnetic fields are by cause of currents produced from the skeletal muscle[2–6]. Over the past four decades, a key challenge has been developing effective methods that offer both high spatial and temporal resolutions. Conventionally, the muscle activity is recorded and analysed electrically with Electromyography (EMG) techniques in the clinic from the skin's using metal or stainless-steel electrodes[7–11]. However, the electric signals suffer from poor spatial resolution when recording EMG signals noninvasively and sensors implanted in the muscle, for more accurate measurements with high-density needle recording probes, face biocompatibility issues. However, in addition to being painful, the penetration of the needle into the muscle disturbs the muscle structure and function. Moreover, in chronic implants, such as for the control of motor prostheses, the interface between the metal contacts of the sensor and the muscle tissue changes over time, leading to infection and rejection by the body. Therefore, a different paradigm enables non-invasive monitoring with the macro-fidelity, temporal, and spatial resolutions and better sensor location and quickly screening without electrical contacts[12,13]. Therefore, the MMG is becoming an alternative method to measure muscle activity and addresses both limitations of the EMG method.

**Fig. 1: Schematic of recording muscle activities with electrically and magnetically techniques.**

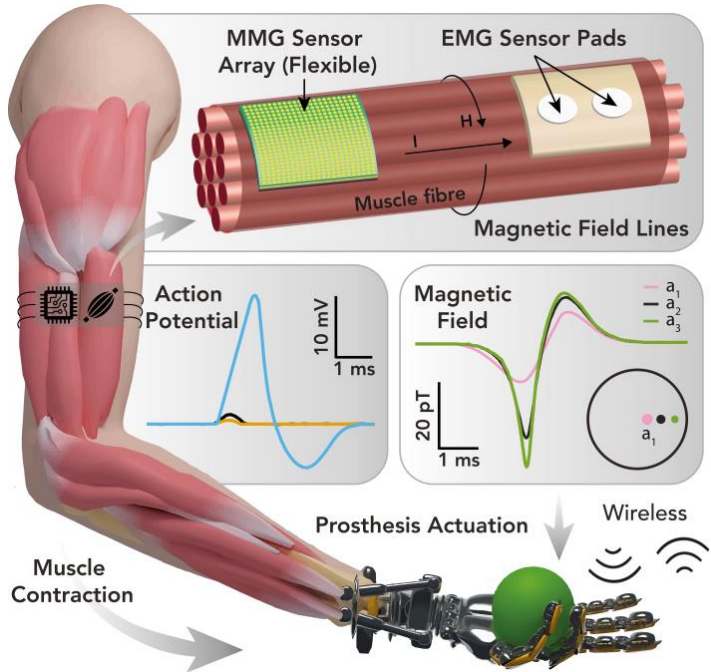

The correspondence between the MMG and EMG methods is defined by the Maxwell-Ampère law, as shown in Fig. 1a. The uncomplicated way of the EMG recording and the similarity of the temporal characteristics between MMG and EMG signals have uplifted the medical and academic society to utilise the EMG technique mainly. The MMG is an alternative technique to measure muscle activities. It addresses both limitations of the EMG means[12]: 1) The magnetic field generated by muscle activity has the same temporal resolution as the EMG signal but offers significantly higher spatial resolution; 2) It does not require electric contacts for recording and hence, the magnetic sensor can be fully encapsulated with biocompatible materials before implantation, minimizing the risk of infection.

Recording such extremely small MMG source amplitudes is still challenging since most of MMG signals are smaller than other active organs and tissues[14]. Fig. 2a illustrates magnetic signals generated from different human body sources with additional environmental interferences. The minimum strength of MMG can reach hundreds of fT/√Hz regarding magnitude spectral density with main frequencies ranging from 10 Hz to 100 Hz, depending on a measurement point to the muscle distance. Therefore, main challenges of MMG measurements stem from the dimension, detection of limit and signal-to-noise ratio of magnetic sensors since the magnitude of geomagnetic fields is roughly five million times bigger, and the background noise can outreach a sub-nT range.

**Fig. 2: Overview of the characteristics of biomagnetic signals and MMG applications.**

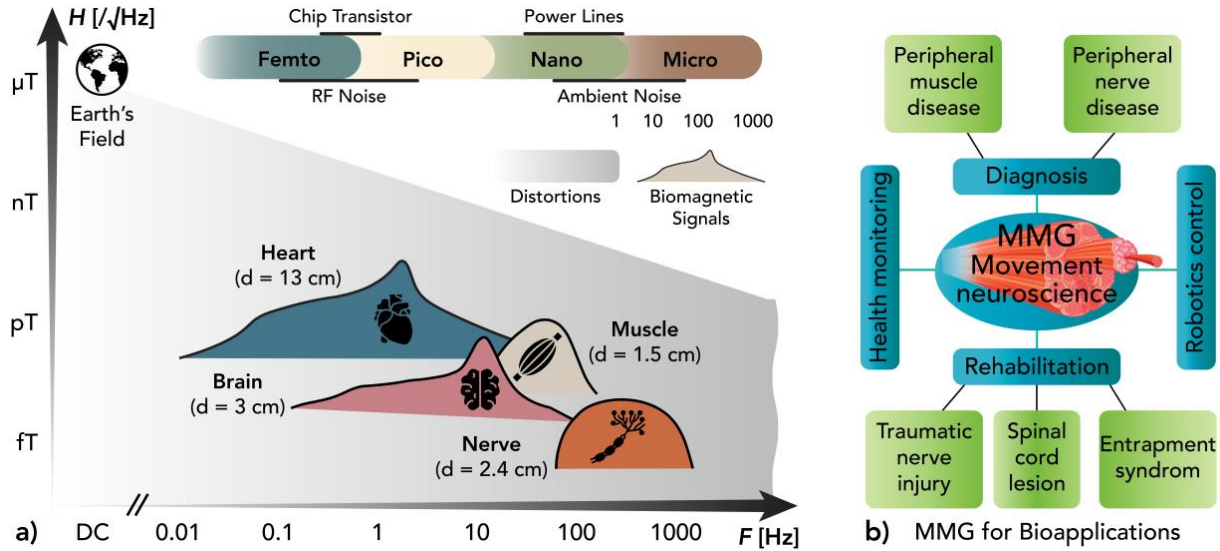

**a** Typical representative biomagnetic signals generated from human active organs and tissues with additionally interferences from the environment[1,12,13,15,16]. **b** Potential application of MMG for diagnosis and rehabilitation of movement disorder, health monitoring and robotics control.

Nowadays, the assessment of MMG signals has become a fundamental benchmark in medical diagnosis[1,17,18], rehabilitation[14,19–21], health monitoring[22–24] and robotics control[25,26], as illustrated in Fig. 2b[12]. Such magnetic information about physiological phenomena is directly associated with human health and wellbeing. The rapid progress of nanotechnology has resulted in a shift towards a remote and continuous recording of diagnosing people's disease on the peripheral muscle and nerve[18]. Each individual needs to understand their physiological and health status by analyzing biomagnetic signals. After that, appropriate treatments can be provided promptly. Triggered by investigating the electrophysiological behaviour of the uterus before delivery, previous MMG in use is chiefly focused on health monitoring during pregnancy[24]. The produced spatial-temporal muscle map will provide a better understanding of the process of labour. Moreover, development of the effective MMG assessment can potentially enhance the quality of life of individuals with sensorimotor conditions such as the traumatic nerve injury, spinal cord lesion and entrapment syndrome[14]. One of the most important MMG research areas is to develop rehabilitation robotics where human-machine interfaces assist the disabled with different limbs to carry out necessary activities of daily life. Currently, the widely and practically used hand prosthesis feed-forward control is only driven by EMG signals, sensing changes of electric potentials at the skin surface from the stump of an amputee due to muscle contractions. This will allow them to manipulate the prosthesis via bending their muscles. However, for the problem of feed-forward control of hand prostheses, the EMG is far from achieving an optimal solution due to the lack of spatial resolution[27]. The MMG becomes an efficient and robust alternative[12,25], for upper-limb prosthesis control, enabling algorithms to extract features from the muscle that can compactly and efficiently depict movement information.

This paper presents the magnetic signals generated by typical skeletal muscles, compare it with electric signals and the relationship between physics and mathematics. The compact muscle model is created and modelled to present context about the magnitude of the MMG signal. The simulation outcomes can serve as the basis for an improved pathophysiological understanding of human skeletal muscles. This would help to revitalize the biomagnetic measurements and constitute an important step to transform the diagnosis of peripheral muscle and nerve diseases and to radically enhance the efficacy of motor rehabilitation after stroke, spinal cord injury or limb loss.

## Results

### 1. Finite-Element Simulation Results

The strength of MMG signals depends on diverse variables. For example, their magnitude can vary from a sub-nT range, when the measurement point is just on the isolated muscle fibres and below the skin, to a sub-pT range, when recording the MMG signals on the outside of the human body[2–4,6]. This section describes the calculation the magnetic field created from a time-changing action potential propagating in a group of skeletal muscle fibres. The position-dependent and the behaviour of the magnetic field on account of the different currents are also performed in the temporal, spectral and spatial domains.

**Fig. 3: Simulated MMG signals by COMSOL with the finite-element method.**

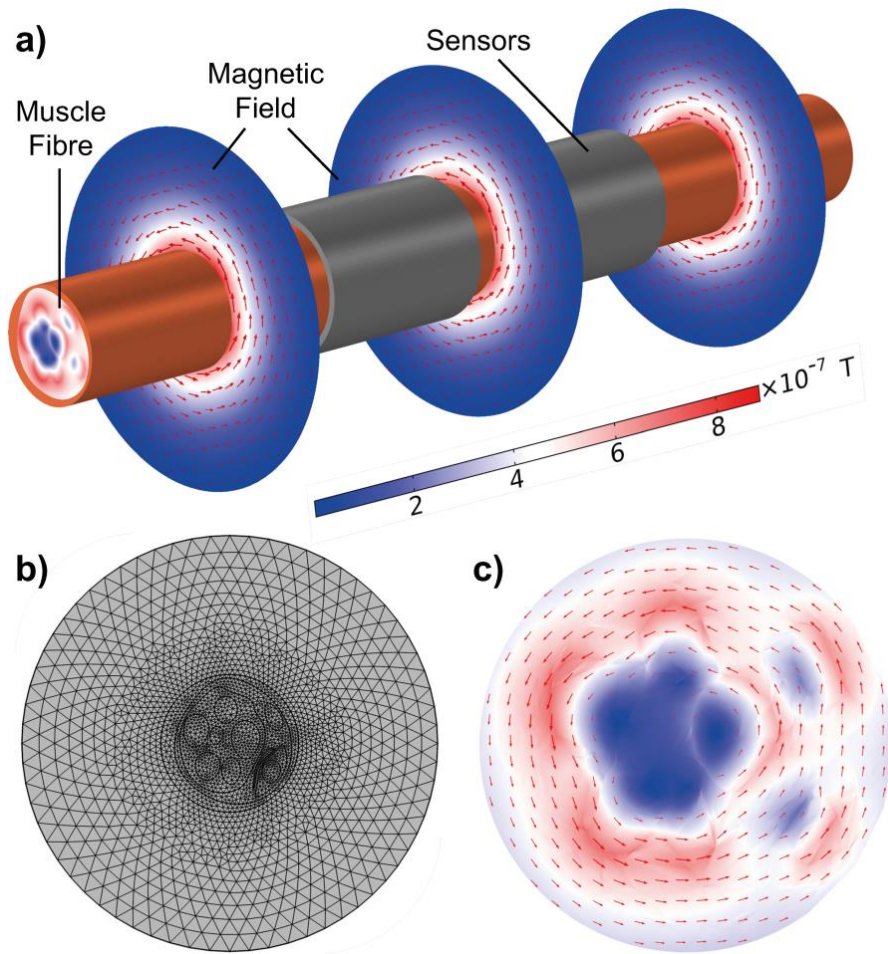

**a** Numerical simulation results of MMG signals in a 3D dimension based on the proposed compact muscle model. **b** computational mesh created in COMSOL software. **c** Results in a 2D dimension with muscle fiber and myofibrils. The red arrows and colour legend indicate the direction and magnitude of the magnetic signal in nano to pico-Tesla range.

We developed the previous approach proposed by Roth and Wikswo[28] to calculate the magnetic field generated by the action potential propagating in skeletal muscle fibres. Due to the presence of currents including fibres, bundles, sheaths of connective tissue, and bath in each area of the system, the contribution to the MMG signal can be eliminated. We extend it to a compact muscle model comprised several fibres, based on the tutorial of a computational framework for peripheral neural interfaces[29] and whose calculation parameters are summarized in Table 1[30,31]. The simulations were performed by Sim4Life (https://zmt.swiss/sim4life/), the open-source PyPNS (https://github.com/chlubba/PyPNS)[32] and COMSOL Multiphysics®, based on the use of finite-element method and detailed neuron models. The simulation results are shown in Fig. 3a with the computational mesh in Fig. 3b. They reveal the effect of the distance between the sensor and a single fibre to investigate the effect of the position of the muscle fibre bundles and radial and axial conductivities of the muscle fibres on the magnitude of the MMG signals. The red arrows and colour legend indicate the direction and magnitude of magnetic fields, respectively. From cross-section result in Fig. 3c, the entire ring magnetic field is the superposition of each muscle fibre with a generated nano to pico-Tesla range.

**Table 1 Described Calculation Parameters**

| Parameter | Description | Value |
| --- | --- | --- |
| $a$ | Fiber radius | $4.00 \times 10^{-5}$ m |
| $b$ | Muscle bundle radius | $1.50 \times 10^{-4}$ m |
| $c$ | Radius of the MB including the sheath | $1.60 \times 10^{-4}$ m |
| $d$ | Distance between fiber and MB centre | $8.00 \times 10^{-5}$ m |
| $\rho_1/\rho_2$ | Field-point radius from the centre of the MB | - |
| $\theta/\theta'$ | Direction of the field point from the centre of the MB | - |
| $\delta$ | Sheath thickness around the MB | $1.00 \times 10^{-5}$ m |

| $U$ | Conduction velocity of action potential | 3.00 m/s |
|---|---|---|
| $\sigma_i$ | Internal conductivity of axons in MB | 0.88 $\Omega^{-1}m^{-1}$ |
| $\sigma_s$ | Conductivity of sheath around MB | 2.00 $\Omega^{-1}m^{-1}$ |
| $\sigma_z$ | Axial conductivity of MB | 5.00 $\Omega^{-1}m^{-1}$ |
| $\sigma_\rho$ | Radial conductivity of MB | variable |

The theoretical and computational modelling were referred to NEURON[33]. The full expressions of these boundary conditions[28] were resolved through MATLAB for linear scalar equations. In addition, we optimized the parameters for the characterization of different currents, duplicating the shape of action potentials measured in the soleus skeletal muscle cells under conditions of the floating electrode technique. Then, the simulation result of transmembrane action potential propagating in muscle fibres is described in Fig. 4a. Subsequently, the *x*, *y* and *z* components of the magnetic field are observed at a point *P*, outside the muscle bundle.

The net magnetic field is illustrated in Fig. 4b, which is obtained from a single muscle fibre at a recording distance *d* from the centre of the bundle. We investigated the relationship between the magnetic fields as a result of different currents, and ratios of axial and radial conductivities of the muscle bundle defined as $\sigma_z$ and $\sigma_\rho$. The shielding effect becomes more significant as the ratio $\sigma_z/\sigma_\rho$ goes up, and therefore the amplitude of the magnetic signal is dropped. That is to say, the bundle current will shield the produced magnetic field if the fibre is near to the bundle centre. At last, the total magnetic field, $B_{total}$, calculated at a distance of 30 $\mu$m from the surface of skeletal muscle at a point *P* with different ratios of $\sigma_z/\sigma_\rho$, is shown in Fig. 4c. It involves four main contributions, the intracellular current for $B_i$, the currents streaming in the bundle for $B_b$, in the sheath for $B_s$ and the external saline for $B_e$. It is worth mentioning that the contribution of saline and sheath currents is much smaller than that of the extracellular bundle currents that hence can be regarded as the main source of shielding[2–4,6].

**Fig. 4: Simulated transmembrane action potential and magnetic signals from the muscle fibre.**

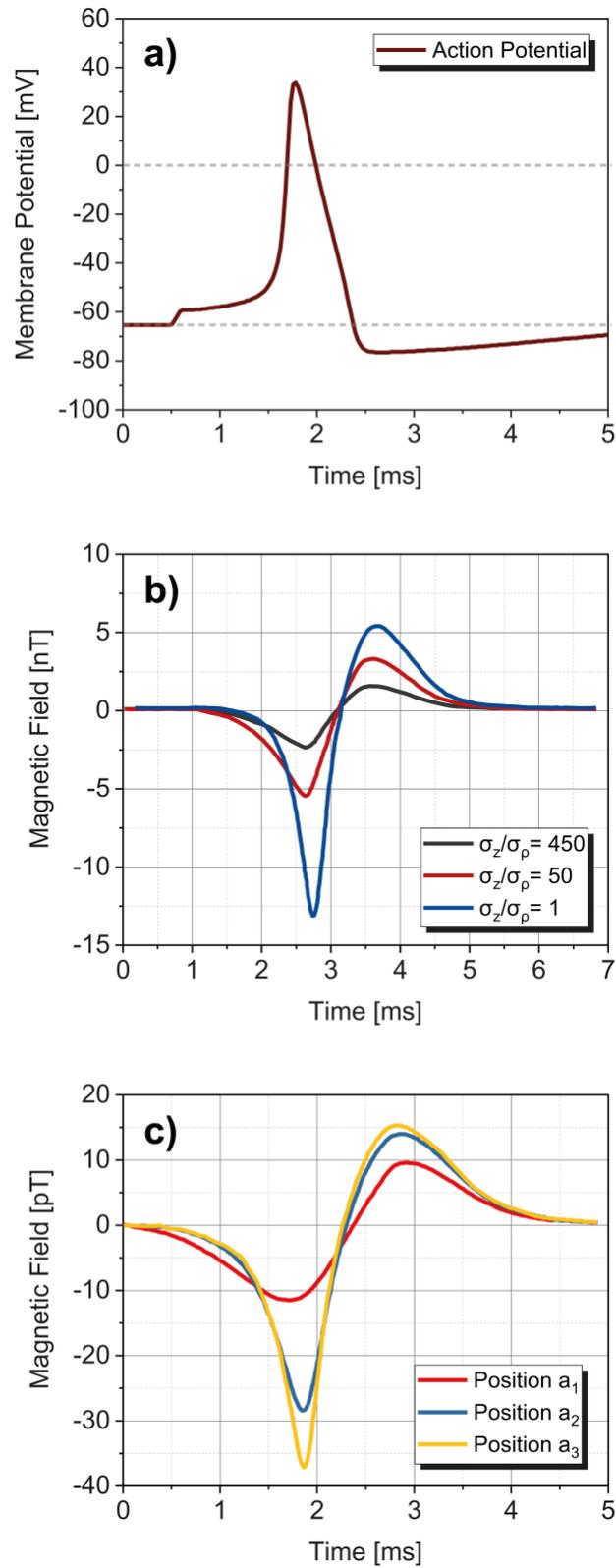

**a** The transmembrane action potential propagating in the muscle fiber. **b** Simulated net magnetic signals at a fixed-point *P*. The magnitude of the MMG signal is a function of distance *d* between the muscle fibre and the recording site *P* with three locations ($a_1$, $a_2$, $a_3$) inside the bundle. This net field involves four main magnetic components, the fiber $B_i$, the bundle $B_b$, the sheath $B_s$ and the saline $B_e$. **c** Total magnetic signals calculated at a distance of 30 $\mu$m from the surface of skeletal muscle resulted from different ratios of $\sigma_z/\sigma_\rho$ ($\sigma_z = 5\ \Omega^{-1}\mathrm{m}^{-1}$). Here, $\sigma_z/\sigma_\rho$ represent axial and radial conductivities.

## 2. Experiment Validation

In 1972, Cohen and Givler discovered the MMG signals using superconducting quantum interference device (SQUID)[2]. Recently, optically-pumped magnetometers (OPM), from competing manufacturers e.g. QuSpin Inc., FieldLine Inc. and Twinleaf with a below 100 fT/√Hz sensitivity[34,35], was utilized to record evoked MMG signals to study the innervation of the nerves in hand muscles in man[16,20]. It has implemented to analyze the signal conduction in muscular fibres, the spatio-temporal dynamics of the magnetic field generated by the propagating muscle action potential[36,37]. However, both experiments, 36 years apart, were conducted in heavily-shielded rooms due to small magnitudes of the MMG signal that can be affected by the magnetic noise in the environment. Spintronics is the study of a fundamental property of electrons known as their spin. Some materials exhibit spin-related magneto-resistive properties at room temperature. That is, their electrical resistance is a function of the magnitude and the direction of the applied magnetic field. This phenomenon has led to the development of spintronic sensor[38,39], which have the potential to detect pico-Tesla level magnetic fields, appropriate for MMG sensing. The size of a typical spintronic sensor is significantly smaller than that of a SQUID or an OPM. Recently, giant magnetoresistance (GMR) sensors were used to record the MMG signal from the surface of a muscle in mice[6]. However, the sensitivity of GMR sensors is in the nano-Tesla range and thus averaging was required to enhance the signal to noise ratio (SNR). Over the last decade, sensing pico-Tesla/√Hz magnetic field has become possible with the tunnelling magnetoresistive (TMR) sensors.

We recently showed, for the first time, identification, characterization and quantification of the MMG signals at room temperature by utilizing a highly miniaturized and sensitive TMR sensors[40]. The sensor array was precisely placed on the hand skin of the abductor pollicis brevis muscle to record the lateral component of the magnetic signal at the room temperature. In the meanwhile, we measured the surface EMG signals at the same location associated with the MMG. Both recordings were carried out in a magnetically shielded environment. As a more visual analysis, the measured MMG signals in the time-frequency domains are plotted as a waterfall, as illustrated in Fig. 5[41]. There was a significant difference in a time-domain between relaxed and strained hands. During a contract moment, an amplitude of 200 pT magnetic fields was continuously generated, which is well-matched with our previous numerical simulation results. However, in the course of a relaxed state, the magnetic signal was approximately 20 to 30 pT, indicating the noise signals of the muscle movement. This noise level also determines the limit of detection of the sensor. To remove the movement artefact and magnetic background noise, a bandpass Butterworth filter (30 to 300 Hz) was utilized. Simultaneously, both peak positions of the EMG and MMG were very close to each other. The observed magnetic signals also approached 200 pT that corresponds with the state-of-the-art experimental outcome from SQUIDs[14]. From MMG signals in a spectral frequency domain with a broadband range, it is noted that the magnitude of the MMG signal at a strained muscle condition is many times larger than the noise level. The signal-to-noise ratio is over 20 among all the bandpass frequency.

**Fig. 5: Analysis of measured MMG signals in a time-frequency domain as a waterfall spectrogram plot using a short-time Fourier transform.**

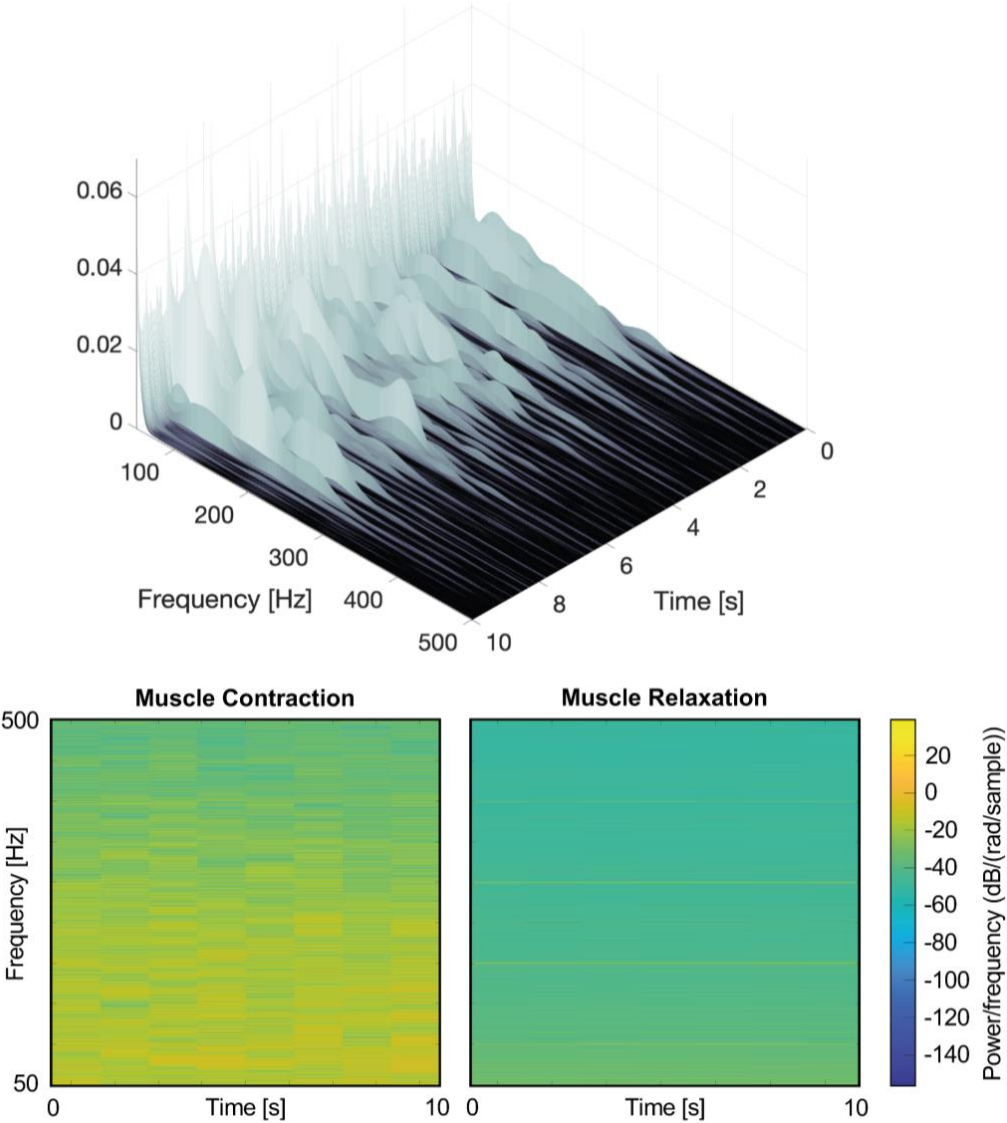

## Discussion

We performed a complete compact model for MMG generation from muscle fibres and motor units is developed and simulated through theoretical analysis. We provide a framework for the interpretation of the MMG signal, similar to knowledge acquired for the EMG signals. The model is based on the Ampere's law, proposed by Roth and Wikswo for the simulation of MMG generated by an axonal fibre. We calculated the magnetic field created from a time-changing action potential propagating in a group of skeletal muscle fibres. That is combined with the mathematical characterization of the generation and extinction of the action potentials at the neuromuscular junctions and tendons, respectively. Moreover, the position-dependent and the behaviour of the magnetic field on account of the different currents in muscle fibres are performed in the temporal, spectral and spatial domains. The simulation outcomes can serve as the basis for a better pathophysiological understanding of skeletal muscles. This will help to revitalize the biomagnetic measurements and constitute a crucial step toward movement neuroscience, diagnosis of peripheral muscle and nerve diseases, and motor rehabilitation. A paradigm-shifting engineering technology would be proposed by interfacing cutting edge theoretical, computational and experimental physics with advanced biomedical modelling and testing. Future descriptions of the action potentials and muscle fibre current density in the spatio-temporal frequency domain will improve this compact model. It will provide an analytical formulation of the magnetic field in the spatial, temporal and respective spectral domains. All these areas can keep rapid numerical calculations and analysis of the MMG signal characteristics related to the underlying physiology, such as the relationship between MMG bandwidth and action potential propagation velocity along the muscle fibres.

## Methods

Theoretical and computational modelling plays an essential role before experimental physics and testing. Here, the strength of MMG signals depends on diverse variables. For example, their magnitude can vary from a sub-nT range, when the measurement point is just on the isolated muscle fibres and below the skin, to a sub-pT range, when recording the MMG signals on the outside of the human body[2–4,6]. In the section, we will provide a compact muscle model to investigate the outcome of the distance between a measurement point and a group of muscle fibres for the magnitude of the MMG signals, depending on different radial and axial conductivities of the muscle bundle. The compact model is also referred to nerve conduction in healthy and pathological motor nerve[42]. The magnetic signal generated from muscle fascicle propagating with a time-changing action potential is illustrated in Fig. 6a, while the detail geometry of the proposed compact muscle model is shown in Fig 6b as a comparison. This is referred and improved the calculation method that was proposed by Roth and Wikswo in 1985[28]. They advocated the use of Ampère's law with its superiority. The magnetic field can be resolved by the contribution of magnetic components from the currents present in each area comprising fibres, bundles, sheaths of connective tissue and bath. We generalized this compact model in regard to a skeletal muscle with a group of fibres. The mathematical calculations and finite-element method (FEM) simulations are performed with MATLAB and COMSOL Multiphysics®. Table I summarize calculation parameters that describe this compact model[30,31]. Before the theoretical demonstration, it is necessary to understand how skeletal muscle electrophysiology works and what happens during a skeletal muscle contraction.

**Fig. 6: Proposed compact muscle model with based on its electrophysiology.**

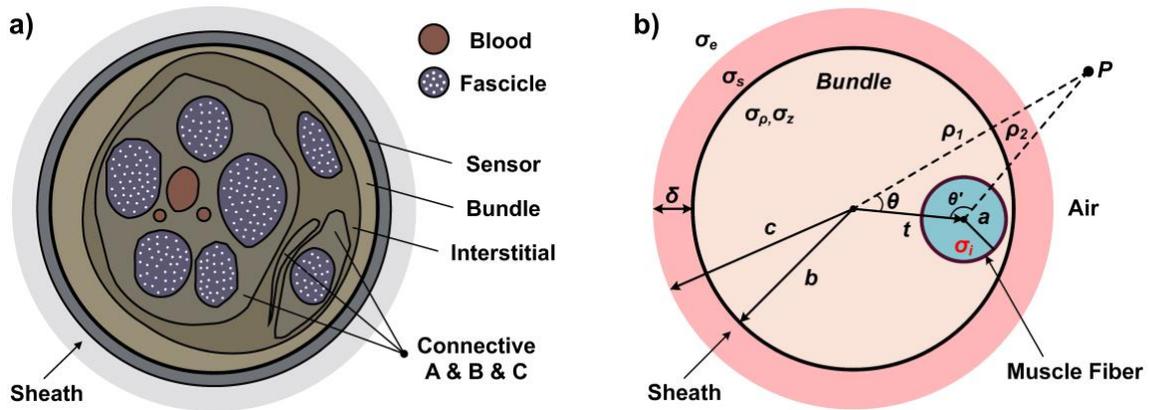

**a** Structure of the skeletal muscle with its electrophysiology. **b** Geometry of the compact muscle model.

# 1. Skeletal Muscle Electrophysiology

Axons and muscle fibres are similar to excitable cables where a large number of $Na^+$ inside flows via voltage-gated $Na^+$ channels. There is an increase in the membrane potential due to part of these electric charges. It is worth mentioning that the other part will stream by way of passive and active $K^+$ and $Cl^-$ conductance. A loop is then formed with source currents from the extracellular space, as illustrated in Fig. 7a. Here, the membrane potential represents a level of capacitance charges, commonly recorded and evaluated through an intracellular microelectrode indicated in Fig. 7a with a blue circle. There exists a membrane depolarization partially as a result of the charge leakage on the excitable cable. Here, an orange ellipse shows inward and outward transmembrane currents through the local influence of the external potential recorded accurately through an extracellular microelectrode. In the front and the back of the active area specified with a grey ellipse in Fig. 7a, the intra-cytoplasmic ions flowing onward the internal potential gradient form axial currents that cannot be directly measured through the traditional electrophysiological approach.

**Fig. 7: Electrical physiology of the muscle fiber as an excitable cable.**

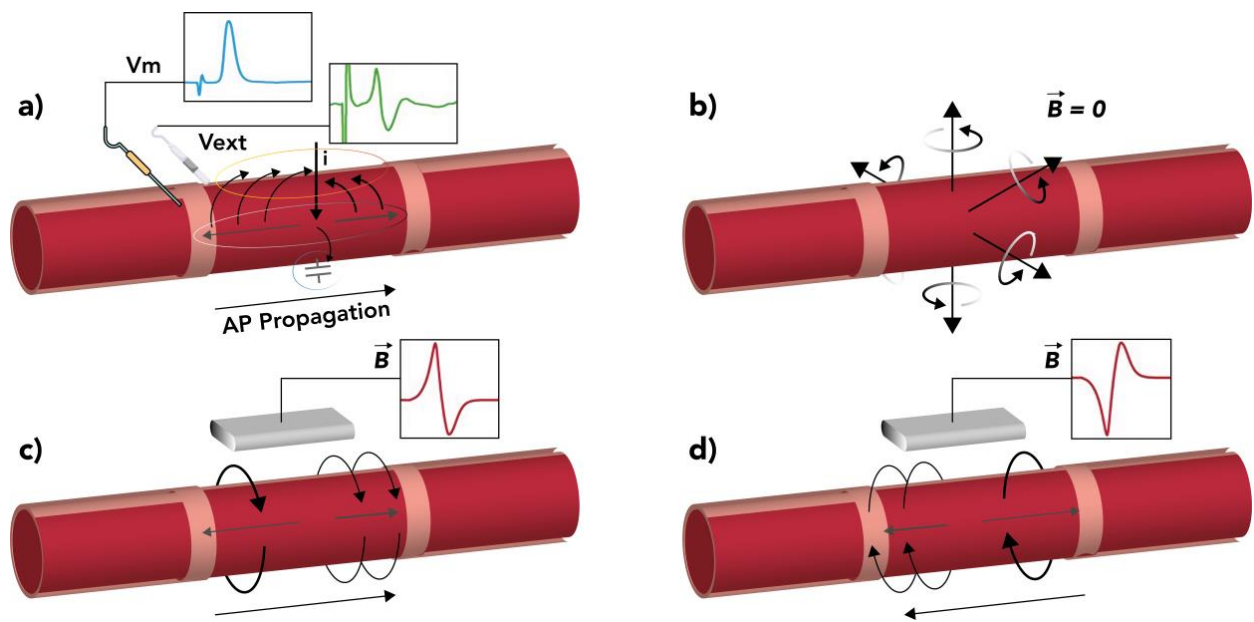

**a** the axial and transmembrane currents flowing at an active area. *Blue ellipse* represents a capacitance of the membrane charges, which is commonly measured by an intracellular pipette; *Orange ellipse* depicts the influence of transmembrane current at an external potential which is recorded through an extracellular electrode; *Gray ellipse* exhibits axial currents, but it should be noted that it cannot be directly measured by the traditional electrophysiological approach. **b** The magnetic fields related to transmembrane currents in theory, which is typically spread along the circumference of a cable. The net magnetic field is nearly zero from a certain distance since magnetic components from each direction of transmembrane currents are nullified with each other; The total magnetic field of the excitable cable related to the axial current in theory where action potential flowing in **c** forward direction and **d** backward direction. This will produce positive and negative magnetic fields, respectively.

In agreement with the electromagnetic theory, any electric current will produce a magnetic field. However, transmembrane currents are not considered as the main magnetic contribution[43,44], as the net magnetic field recorded from a long distance compared with a diameter of the excitable cable is close to zero, shown in Fig. 7b. The uniform distribution of the circumferential ion channels along the neurites, while axons and cell bodies will nullify magnetic components from each direction of transmembrane currents. Therefore, it is expected that only axial currents offer a benefaction to the net magnetic field, as illustrated in Fig. 7c and Fig. 7d. The axial current directions determine the generated positive and negative magnetic signals. Besides, there is another contribution from the current reflux streaming in the extracellular volume. It, thus, partially and oppositely filters the intracellular source.

Before further understanding of the magnetic quantity, the associated bioelectricity of the neuromuscular system should be investigated firstly since the magnetic signal depends on the shape and propagation speed of the action potential. Nowadays, electrical recording of the muscle activity has been a well-established method over decades in academic research and medical diagnostic. An electrophysiological characterization is commonly carried out under a specific experiment environment in nerve-muscle of rodent animals, such as small rats. Conventionally, intracellular measurements in the skeletal muscle of rats cannot be easily performed because of macroscopic movements during contraction. Benefiting from a floating electrode strategy[45,46], this problem was resolved, finally maintaining a continuous measurement within the moving organs and tissues, as demonstrated in Fig. 8, in this way, to understand the underlying mechanism of the human skeletal muscle. In the next section, we will discuss and compare the magnetic signals with electric signals and the relationship between physics and mathematics. In a real measurement process, the MMG signals are achieved by averaging signals using the EMG signals as a trigger.

**Fig. 8: Electrical physiology of the muscle fiber as an excitable cable.**

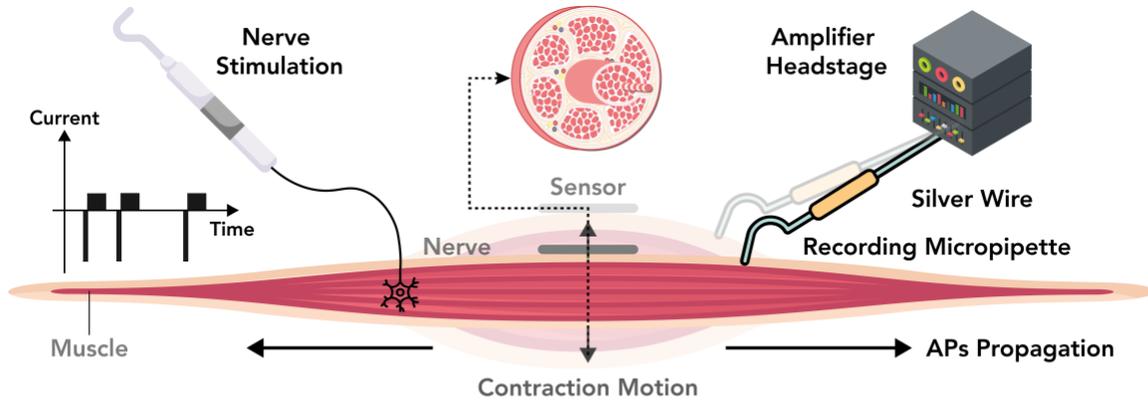

## 2. Single Fibre Model

**Precondition:** The applied current volume source element, $\bar{J}^i$, is distributed in a finite and inhomogeneous conductor.

The current density on the entire volume conductor will generate a magnetic field, expressed as[47]

$$4\pi\bar{H} = \int_0^v \bar{J} \times \nabla\left(\frac{1}{r}\right) dv \tag{1}$$

where $r$ is the distance from an external field point at which is evaluated to an element of volume $dv$ inside the body, $\bar{J}dv$ is a source element and $\nabla$ is an operator with respect to the source coordinates. Substituting below equation (2),

$$\bar{J} = -\sigma \nabla \Phi + \bar{J}^i \tag{2}$$

into equation (3) and dividing the inhomogeneous volume conductor into homogeneous regions $v_j$vj with conductivity $\sigma_j$, it can be obtained that

$$4\pi \bar{H} = \int_0^v \bar{J} \times \nabla\left(\frac{1}{r}\right) dv - \sum_v \int_0^{v_j} \sigma_j \nabla \Phi \times \nabla\left(\frac{1}{r}\right) dv \tag{3}$$

If we use the vector identity $\nabla \times \Phi \bar{A} = \Phi \nabla \times \bar{A} + \nabla \Phi \times \bar{A}$, subsequently the integrand of the last term of equation (3) can be expressed as $\nabla \times [\Phi \nabla(1/r) - \Phi \nabla \times \nabla(1/r)]$. As $\nabla \times \nabla \Phi = 0$ for any $\Phi$, the last term involving its sign can be replaced as

$$-\sum_j \int_0^{v_j} \sigma_j \nabla \times \Phi \nabla\left(\frac{1}{r}\right) dv \tag{4}$$

Regarding the vector identity, the following equation is adopted from Page 604[47]:

$$\int_0^v \nabla \times \bar{C} dv = -\int_0^S \bar{C} \times d\bar{S} \tag{5}$$

where the surface integral is taken over the surface $S$ bounding the volume $v$ of the integral. By applying equation (5) to equation (4), the last term in equation (3), including its sign, can now be replaced by

$$\sum_j \int_0^{S_j} \sigma_j \Phi \nabla\left(\frac{1}{r}\right) \times d\bar{S}_j \tag{6}$$

Finally, applying this result to equation (3) and again specifying the conductive primed area and the double-primed area inside and outside the boundary, respectively, and then adjusting $d\bar{S}_j$ from the primed to the double-primed area, we obtain (Each interface appears twice, once as the surface of $v_j$ and followed by the surface of each adjacent area of $v_j$)

$$4\pi \bar{H}(r) = \int_0^v \bar{J}^i \times \nabla\left(\frac{1}{r}\right) dv + \sum_j \int_0^{S_j} (\sigma_j'' - \sigma_j') \Phi \nabla\left(\frac{1}{r}\right) \times d\bar{S}_j \tag{7}$$

Here, the equation (7) explains the external magnetic field of a finite conductor including an internal electric volume source $\bar{j}^i$ and an inhomogeneity $(\sigma_j'' - \sigma_j')$, which has been first proposed by David Geselowitz in 1970[48]. It is worth mentioning that the first term on the right side of equation (7), containing $\bar{j}^i$, indicates the contribution of the volume source, whilst the second term denotes the influence of the boundary and inhomogeneity. Virtually, the source $\bar{j}^i$ comes from cellular activities, thus, offering diagnostic value. Regarding the second term, there is a deformation caused by an unevenness of the volume conductor.

When evaluating the generated electric field generated, these sources are very similar. As equation (8) below in the electrical case, there exists both primary and secondary sources as following

$$4\pi\sigma\Phi(r) = \int_0^v \bar{j}^i \cdot \nabla\left(\frac{1}{r}\right)dv = + \sum_j \int_0^{S_j} (\sigma_j'' - \sigma_j')\Phi\bar{n}_j \cdot \nabla\left(\frac{1}{r}\right)d\bar{S}_j \tag{8}$$

By the same token, it is simple to concede, from the conjunction with equation (8), that if we consider a homogeneous volume conductor, the value $\sigma_j'' - \sigma_j'$ in the second term thus is identical to zero and this second expression disappears. Eventually, on account of this homogeneous distribution of the volume source in the finite conductor, the equation (7) is simplified to the magnetic field equation, a magnetic dipole moment of a current volume distribution in connection with any origin, which can be expressed as

$$\bar{m} = \frac{1}{2}\int_0^v \bar{r} \times \bar{j}^i dv - \frac{1}{2}\int_0^v \bar{r} \times \sigma\nabla\Phi dv \tag{9}$$

For precise MMG measurements with high spatial and temporal resolutions, the target is to eliminate the influence of secondary sources as much as possible. Finally, through observing from equation (7), we can conclude that the conductivity discontinuity is equal to the secondary surface source $\bar{K}_j$ that is defined as $\bar{K}_j = (\sigma_j'' - \sigma_j')\Phi\bar{n}$ in which $\Phi$ represents a surface potential on $S_j$. Here, the $\bar{K}_j$ depicts an identical secondary current source as the magnetic field, inspired by the electric field in equation (8).

## 3. Calculation of Magnetic Lead Field

In 1972, the reciprocity theory was expanded to the time-varying biomagnetism based on the magnetic signal from the human heart[49]. The subsequent development goes with a route that is similar to the proof of a reciprocity theory. Benefiting from this theorem, the equations for the magnetic measurement can be derived. In the following discussion, the subscript L in equations represents *lead* whilst there is another subscript M indicating *magnetic leads* on account of the reciprocal current of the derivative per unit time. Here, an induced current in the conductor is determined by the variation rate of the magnetic flux connecting a current loop. The time-varying electrified mutual current, $I_r$, is normalized, similar to the electric field situation. Thus, the time derivative of the current is uniform for all $\omega$ values. We can easily derive essential equations in regard to the magnetic recording based on the foundation of corresponding electric formulas.

The basic bipolar lead of the biomagnetic recording mainly consists of a solenoid coil whose magnetic core and disc-shaped terminals have infinite permeability, as demonstrated in Fig. 6. If we set up an electric current, a magnetic field will be established due to equal and opposite charges at the core terminals. The word *magnode* was firstly proposed by Baule and McFee in 1963[50] after Michael Faraday came up with the word *electrode* in 1834. Standing on the shoulders of the pioneers, the bipolar magnetic lead was created and very similar to bipolar electric lead. When mutual inductance current is fed into the basic magnetic lead, a scalar reciprocal magnetic potential field, $\Phi_{LM}$, is generated in an infinite space with uniform permeability. It has the same spatial characteristic as $\Phi_{LE}$ that depicts the scalar reciprocal electric potential field in an infinite space with uniform conductivity. The *electrodes* are placed at positions associated with the *magnodes*. It is noted that both $\Phi_{LE}$ and $\Phi_{LM}$ will be uniform in the central area when the *electrodes* and *magnodes* keep parallel, and both have a more significant volume compared with their spacing.

We have summarized equations of the lead field in theory used for electric and magnetic recordings in Table 2. Both electric and magnetic scalar potentials, $\Phi_{LE}$ and $\Phi_{LM}$, are spatially dependent and can be derived from the Laplace's equation. They enjoy identical forms with regard to similar shape and position of the *electrode* and *magnode* without the influence of the unevenness of the volume conductor or the boundary with the air. Besides, the magnetic and electric signals, $V_{LM}$ and $V_{LE}$, also have the same shape. However, there is a difference of sensitivity distributions between magnetic and electric measurements of the applied current density, $\bar{J}^i$. This is because of the different forms, the curl of the reciprocal magnetic lead field, LM, and reciprocal electric lead field, LE.

**Table 2 Theoretical equations in electric and magnetic lead fields.**

| Description | Electric Lead | Magnetic Lead |
| --- | --- | --- |
| Field is a negative gradient of the scalar potential in a reciprocating excitation way | $\bar{E}_{LE} = -\nabla \Phi_{LE}$ | $\bar{H}_{LM} = -\nabla \Phi_{LM}$ |
| Magnetic induction due to reciprocal energization | - | $\bar{B}_{LM} = \mu \bar{H}_{LM}$ |
| Reciprocal electric field | $\bar{E}_{LE} = -\nabla \Phi_{LE}$ | $\bar{E}_{LM} = ''\bar{r} \times \bar{B}_{LM}$ |
| Lead field (current field) | $\bar{J}_{LE} = \sigma \bar{E}_{LE}$ | $\bar{J}_{LM} = \sigma \bar{E}_{LM}$ |
| Detected signal when: $I_{RE} = 1\ A$, $dI_{RE}/dt = 1\ A/s$ | $V_{LE} = \int \frac{1}{\sigma} \bar{J}_{LE} \cdot \bar{J}^i dv$ | $V_{LM} = \int_0^v \frac{1}{\sigma} \bar{J}_{LM} \cdot \bar{J}^i dv$ |

## 4. The Source of the Magnetic Field

An unbounded homogeneous medium is required for the conductivity to be dual to the magnetic permeability, where the latter is uniform in the body and in space. As in electric measurements, it is possible to create compound magnetic leads by connecting any number of detectors together. We investigate how the nature of the magnetic lead field $\bar{J}_{LM}$ produced by reciprocal energization of the coil of the magnetic detector with a current $I_r$ at an angular frequency. Using the same sign convention between the energizing current and the measured voltage as in the electric case, we obtain the corresponding situation for magnetic measurements, as illustrated in Fig. 9.

This section provides an alternative description of the source of the magnetic field sensed by magnetic pickup coils (which is valid for the case of axial symmetry). By substituting the reciprocal electric field, $\bar{E}_{LM} = -(\mu \bar{r}/2) \times \nabla \Phi_{LM}$, into the magnetic lead field current density, $\bar{J}_{LM} = \sigma \bar{E}_{LM}$, we obtain the voltage $V_{LM}$ in cylindrical coordinates as

$$V_{LM} = -\frac{\mu}{2}\int_0^v (\bar{r} \times \nabla \Phi_{LM}) \cdot \bar{J}^i dv = \frac{\mu}{2}\int_0^v \nabla \Phi_{LM} \cdot (\bar{r} \times \bar{J}^i) dv \qquad (10)$$

**Fig. 9: Simplified schematic of a bipolar magnetic lead.**

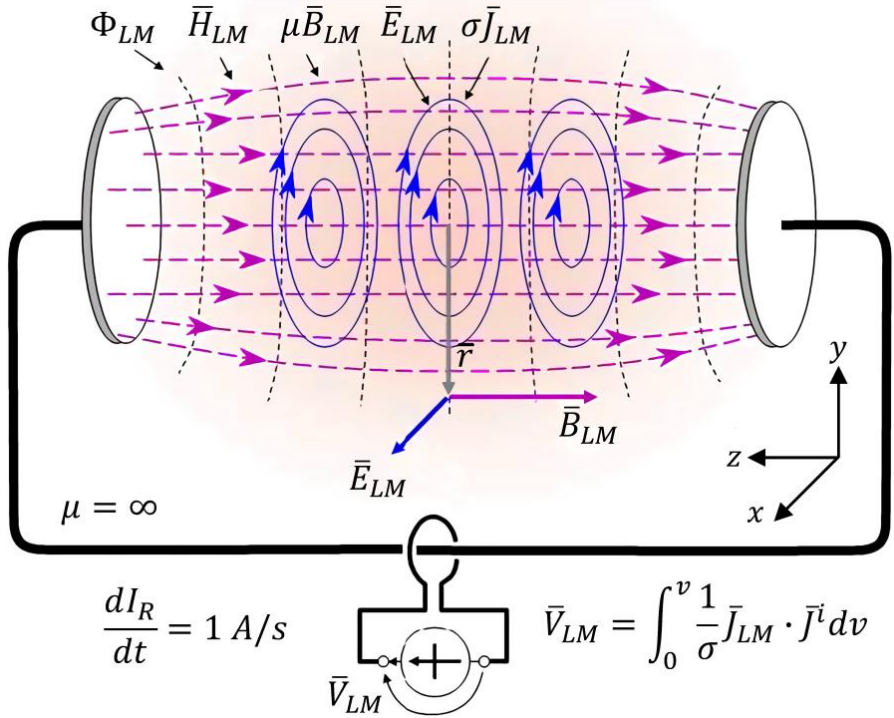

Using the vector identity $\nabla \cdot (\Phi_{LM} \bar{r} \times \bar{J}^i) = \Phi_{LM} \nabla \cdot (\bar{r} \times \bar{J}^i) + \nabla \Phi_{LM} (\bar{r} \times \bar{J}^i)$, we obtain

$$V_{LM} = \frac{\mu}{2}\int_0^v \nabla \cdot (\Phi_{LM} \bar{r} \times \bar{J}^i) dv - \frac{\mu}{2}\int_0^v \Phi_{LM} \nabla \cdot (\bar{r} \times \bar{J}^i) dv \qquad (11)$$

Applying the divergence theorem to the first term on the right-hand side and using a vector expansion, i.e., $\nabla \cdot (\bar{r} \times \bar{j}^i) = \bar{j}^i \cdot \nabla \times \bar{r} - \bar{r} \cdot \nabla \times \bar{j}^i$ on the second term of equation (11), and noting that $\nabla \times \bar{r} = 0$, we obtain

$$V_{LM} = \frac{\mu}{2} \int_0^v \Phi_{LM}(\bar{r} \times \bar{j}^i) d\bar{S} + \frac{\mu}{2} \int_0^v \Phi_{LM} \bar{r} \cdot \nabla \times \bar{j}^i dv \quad (12)$$

Since $\bar{j}^i = 0$ at the boundary of the medium, the surface integral equals zero, and is written as

$$V_{LM} = \frac{\mu}{2} \int_0^v \Phi_{LM} \bar{r} \cdot \nabla \times \bar{j}^i dv \quad (13)$$

where the amount $\Phi_{LM}$ represents the magnetic scalar potential, which is as a result of the reciprocating energization of the pickup wire. The *vortex source*, $\bar{I}_v$, then has the form

$$\bar{I}_v = \nabla \times \bar{j}^i \quad (14)$$

This is the magnetic source strength in equation (13). It is noted that this *vortex* is assigned a definition derived from *curl* that represents the circulation per unit area, described as:

$$\nabla \times \bar{A} = \frac{1}{\Delta S} \oint_0^{\Delta S} \bar{A} \cdot d\bar{l} \quad (15)$$

It depicts that we can obtain the line integral in the region of $\Delta S$ at any point to make it oriented in the field and ultimately boost the integral at a direction of a curl.

## 5. Characterize Soleus Skeletal Muscle Currents

Results obtained from the myocyte voltage-clamp recordings are combined with recordings of synaptic conductance and an excitatory postsynaptic potential (EPSP) in the flexor digitorum brevis (FDB) of the mouse, to reproduce the action potential measured in the soleus skeletal muscle. The analysis process is described as follows:

***Synaptic Current***. The synaptic current was modelled as a current of the form: $I_{syn} = g_{syn}(t) \cdot (V - E_{syn})$, with reverse potential, $E_{syn} = 0$, and synaptic conductance of the form, $g_{syn}(t) = g_{syn,max} \cdot exp(-t/\tau_{syn})$. The synaptic decay time constant was fitted from recordings of synaptic conductance of the FDB under voltage-clamp and set to $\tau_{syn} = 0.58\ ms$. Since the maximal conductance, $g_{syn,max}$, recorded in the FDB was not sufficient to elicit an action potential in our model fibre, we choose to set the maximal synaptic conductance at the minimal value necessary to elicit an action potential, which was $g_{syn,max} = 10\ \mu S$.

***Kir and Leak Currents***. The characteristics of the Kir current were obtained from the myocyte voltage-clamp study. Myocyte Kir current was expressed by the form: $I_{Kir} = g_{max}m(V - E_K)$, where the maximal conductance $g_{max}$ is set to 600 $\mu S/cm^2$, and the variable *m* evolves in time according to:

$$m(t) = m_\infty - (m_\infty - m_0) \cdot e^{\,t/\tau_m} \tag{16}$$

where $m_\infty$ and $\tau_m$ where determined through fitting of the isolated Kir current, as:

$$m_\infty = \frac{1}{1 + exp(-0.074(-91.6 - V))} \tag{17}$$

and $\tau_m = 0.2\ ms$. The leak current was then adjusted to reproduce the decay dynamics of the EPSP measured in FDB voltage-clamp recordings (Fig. 10). The leak current was modelled as: $I_{leak} = g_{leak}(V - E_{leak})$, where $g_{leak} = 200\ \mu S/cm^2$ and $E_{leak} = -90\ mV$.

***Action potential.*** There is a sharp jump with a resulting decrease of voltage on the cell membrane or the membrane potential with a notable feature. Generally, this response process is needed adequate current to initiate the cell membrane. In other words, the action potential will not be triggered if the current does not reach a threshold level for the membrane depolarization. The whole response is divided into five stages. 1) *Stimulation*: First of all, the reaction starts from a sharp jump as a result of stimulation in the action potential. As illustrated in Fig. 10, to initiate membrane depolarization, we should apply an adequate current to arrive and exceed the threshold voltage; 2) *Depolarization*: The rapid increase of the membrane potential will open a sodium channel where a large number of sodium ions pour in; 3) *Repolarization*: The rapid inactivation of the sodium ion channel and activated a potassium ion channel leads to a large amount of potassium ion outflow, which gives rise to a membrane repolarization; 4) *Hyperpolarization*: This process is as a result of the potassium outflow and potassium channel closure, causing a drastic decrease in the membrane potential; 5) *Resting-State*: Ultimately, the membrane potential restores to the initial state before the next stimulation occurs.

**Fig. 10: Electrical physiology of the muscle fiber as an excitable cable.**

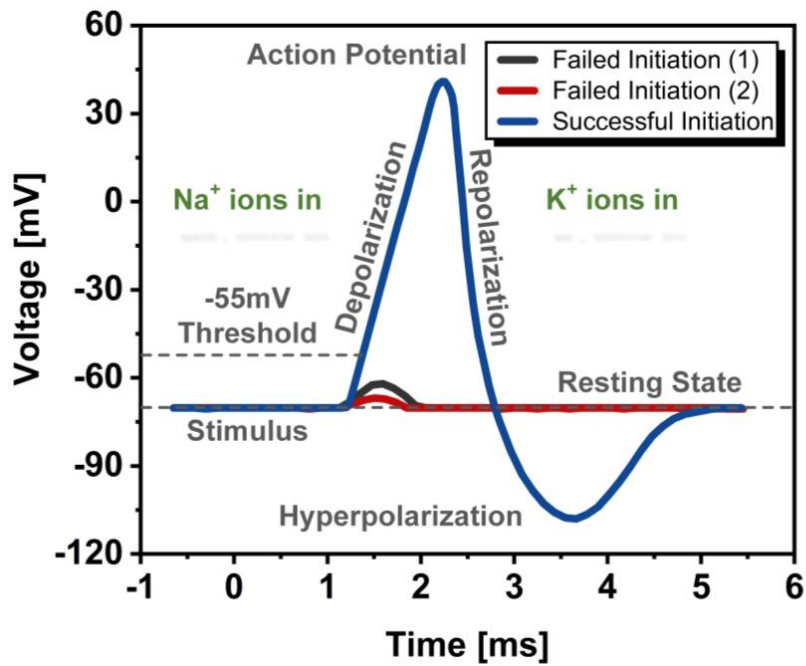

*Sodium and Potassium Currents*. After determining the synaptic, Kir and leak currents parameters, we adjusted the parameters of sodium and potassium currents in order to fit the average action potential recorded during floating electrode recordings on the soleus skeletal muscle. Since the average baseline membrane potential was slightly depolarized due to pipette leak, we added in the compartment corresponding to the recording site, a pipette leak current of conductance $g_{pipette} = 43\ \mu S/cm^2$ and reversal potential $e_{pipette} = 0$. After adding the pipette leak, we adjusted the sodium and potassium currents that were modelled as follows:

$$I_{Na} = g_{Na,max} \cdot m^3 \cdot h(V - E_{Na}) \tag{18}$$

with $g_{Na,max} = 0.028\ S/cm^2$ and $E_{Na} = 50\ mV$ and:

$$I_{Na} = g_{Na,max} \cdot m^3 \cdot h(V - E_{Na}) \tag{19}$$

$$\tau_m = 0.12 \cdot exp\ (-0.01354(V + 55)) \tag{20}$$

$$h_\infty = \frac{1}{1 + exp(-0.8(-50 - V))} \tag{21}$$

$$\tau_h = 0.48 \cdot exp\ (-0.01252(V + 22)) \tag{22}$$

$$I_{(K,TEA)} = g_{(K,TEA,max)} n^4 (V - E_K) \tag{23}$$

with $g_{(K,TEA,max)} = 0.02\ S/cm^2$ and $E_K = -90\ mV$ and:

$$n_{(\infty,TEA)} = \frac{1}{1 + exp(0.06(-30 - V))} \tag{24}$$

$$\tau_{(\infty,TEA)} = 1.6 \cdot exp\ (-0.005(V + 20)) \tag{25}$$

$$I_{(K,4AP)} = g_{(K,4AP,max)} n^4 (V - E_K) \tag{26}$$

with $g_{(K,4AP,max)} = 0.2\ S/cm^2$ and $E_K = -90\ mV$ and

$$n_{(\infty,4AP)} = \frac{1}{1 + exp(0.08(-36 - V))} \tag{27}$$

$$\tau_{(\infty,4AP)} = 1.6 \cdot exp(-0.0193(V - 79)) \tag{28}$$

*Axial Intracellular Currents*. In "Magnetic Recordings" section, we show that transmembrane currents do not generate a detectable magnetic field. This result was not unexpected both since transmembrane currents generate opposite contributions to magnetic fields, thanks to a quasi-homogeneous distribution of channels on the membrane surface and because the magnetic field produced by transmembrane currents is expected to be negligible compared to that of axial currents (Woosley et al. 1985; Hamalainen et al.

1993). On the contrary, currents flowing along the muscle axis were found to be the primary generators of the recorded magnetic field. The axial intracellular currents in each compartment i, were

$$I_i = \frac{(V_{i+1} - V_i)}{R_i} \tag{29}$$

where $V_i$ and $R_i$ are respectively the membrane potential and the axial resistance of compartment $i$. $R_i$ can be expressed in term of specific resistivity $\rho$ as:

$$R_i = \frac{\rho l}{A} \tag{30}$$

where $l$ and $A$ are length and cross section of the compartment.

Even though the transmembrane current does not dedicate to a detectable magnetic signal, they are at the origin of the currents streaming in the extracellular medium[51]. If a local source and sink of current (a depolarizing sodium current and a leak current) are present at two different sites along the cable, local charge variations in the extracellular medium create a potential gradient with opposite polarity with respect to the intracellular space. Hence, extratial and intracellular gradients generate currents in opposite directions. Because of the small extracellular resistivity, extracellular currents are generally smaller than intracellular currents and dispersed, at least in the cortex, in a larger volume. On the contrary, inside of the muscle, the fibres are densely packed, and the extracellular current is restricted to flow along the fibres in a gap of a few microns. Since the axial resistance is inversely proportional to the cross section as in equation (30), this small value means high axial resistance, which in turn translates to larger potential gradients. As shown in "Electrophysiological Recordings", when action potentials were triggered in the muscle, potential peak-to-peak amplitude of around 6 mV was recorded in the extracellular medium. Extracellular currents in the muscle are then likely to contribute considerably to the generation of the magnetic field.

## 6. Magnetic Field from Soleus Skeletal Muscles

The approach developed by Roth and Wikswo[28] is used to calculate the magnetic field generated by the action potential propagating in a single muscle cell. Here, we extend it to a compact muscle model comprised several fibres. For the sake of clarity, we summarize here the main lines of the method, and we describe how the calculations were generalized for the case of the entire muscle. The muscle fibre was modelled as a cylindrical cable composed of 1200 compartments with 10 µm length and 50 µm diameter. A cylindrical fibre of diameter $a = 50$ µm was placed at distance $t$ from the centre of the bundle. The bundle had a diameter $b = 150$ µm and 50 µm diameter fibres separated by a 10 µm interstitial space and was surrounded by a sheath with thickness $\delta = 10$ µm. Two coordinates systems are necessary to discuss the bundle. One is the primed system centred at the fibre, and the other is an unprimed system centred at the bundle. Both are related to simple relationships[28].

The interior of the fibre and the physiological saline had homogeneous, isotropic conductivities $\sigma_i$ and $\sigma_e$. In order to mimic both the presence of the other fibres, that limit the diffusion of the currents along the radial direction and the interstitial space, that, instead, favours the diffusion of the currents along the axial direction in the near surrounding of the fibre surface, the bundle itself was modelled as an anisotropic medium, with different conductivities along the radial and axial directions, $\sigma_\rho$ and $\sigma_z$. The bundle itself was then surrounded by a very thin sheath with conductivity $\sigma_s$.

The calculation of the magnetic field passes first through the calculation of the potential in each of the regions of the system. This can be resolved via Laplace's equation with suitable boundary conditions. Since we know from the simulation of the action potential dynamics, the value of the transmembrane potential along with the fibre at every point in time, the first boundary condition consists in assuming that the potential at $\rho' = a$ is equal to the membrane potential, $\phi_m(z)$, at each point $z$ along with the fibre. Furthermore, the potential has to be continuous at the boundaries, $\rho = b$ and $\rho = c$. It is noted that there are normal components of the current density over these interfaces continuously. The potential in the bundle, which is anisotropic, also solves the Laplace's equation provided the following coordinate transformation: $\rho^* = \sqrt{(\sigma z / \sigma \rho)}\rho$.

The electric potential in the four regions can be written in the Fourier space at a cylindrical coordinate as an extension of the eigenfunction of the Laplace's equation, that is, according to the Fourier summation of the trigonometric functions and modified Bessel functions,

*Fibre:*

$$\phi_i(\rho', \theta', k) = A_0(k)I_0(|k|\rho') + 2\sum_{m=1}^{\infty} A_m(k)I_m(|k|\rho')\cos(m\theta') \tag{31}$$

*Bundle:*

$$\phi_b(\rho'^*, \theta', k) = B_0(k)I_0(|k|\rho'^*) + C_0(k)K_0(|k|\rho'^*) \\ + 2\sum_{n=1}^{\infty}(B_n(k)I_n(|k|\rho') + C_n(k)K_n(|k|\rho'^*))\cos(n\theta') \tag{32}$$

*Sheath:*

$$\phi_s(\rho, \theta, k) = D_0(k)I_0(|k|\rho) + E_0(k)K_0(|k|\rho) \\ + 2\sum_{m=1}^{\infty}(D_m(k)I_m(|k|\rho) + E_m(k)K_m(|k|\rho))\cos(m\theta) \tag{33}$$

*Saline:*

$$\phi_e(\rho, \theta, k) = F_0(k)K_0(|k|\rho) + 2\sum_{m=1}^{\infty} F_m(k)K_m(|k|\rho)\cos(m\theta) \tag{34}$$

where $k$ denotes the spatial frequency and $A_m(k)$, $B_m(k)$, $C_m(k)$, $D_m(k)$, $E_m(k)$ and $F_m(k)$ are unknown coefficients that have to be determined from the boundary conditions.

As mentioned above, there are six boundary conditions: one, at $\rho' = a$, imposes that the Fourier transform of the membrane potentials is identical to the difference of the intracellular and bundle potential, $\phi_m(k) = \phi_i(a, \theta', k) - \phi_b(a, \theta', k)$. The others five conditions are given imposing the continuity of the potential at $\rho = b$ and $\rho = c$ and of the current density component across these surfaces.

The full expressions of these boundary conditions are detailed in [28] and were resolved via MATLAB for linear scalar equations. We used $m = 6$ terms to approximate the infinite series in equation (31) to equation (34).

Once the expressions of the potential are obtained, the current density can be calculated by differentiating equation (31) to equation (34) and through the equality $J = -\sigma\nabla\phi$. From the current density [28], one calculates the azimuth element of the magnetic field averaged over the entire angle $\theta$ using the Ampere's law. The integral on $\theta$ removes all terms in the current density expression, except for the term $m = 0$, while other terms are used as Fourier sums, hence greatly simplifying the calculation. The integration performed on each region of the system represents the expression of the magnetic field in the Fourier space, $B$, as follows:

$$B(\rho, k) = B_i(\rho, k) + B_b(\rho, k) + B_s(\rho, k) + B_e(\rho, k) \tag{35}$$

where

*Fibre:*

$$B_i(\rho, k) = i\frac{\mu_0 \sigma_i k a}{\rho |k|} A_0(k) I_1(k|a|) \tag{36}$$

*Bundle:*

$$B_b(\rho, k) = B_{b1}(\rho, k) + B_{b2}(\rho, k) + B_{b3}(\rho, k) \tag{37}$$

where

$$B_{b1}(\rho, k) = i\frac{\mu_0 \sigma_z k}{\rho |k|} \sqrt{\frac{\sigma_\rho}{\sigma_z}} \left[ \left[ B_0(k) I_0\left(|k|\sqrt{\frac{\sigma_z}{\sigma_\rho}} t\right) + 2\sum_{n=1}^{\infty} B_n(k) I_n\left(|k|\sqrt{\frac{\sigma_z}{\sigma_\rho}} t\right) \right] \right.$$

$$\left. + \left[ C_0(k) K_0\left(|k|\sqrt{\frac{\sigma_z}{\sigma_\rho}} t\right) + 2\sum_{n=1}^{\infty} C_n(k) K_n\left(|k|\sqrt{\frac{\sigma_z}{\sigma_\rho}} t\right) \right] \cdot t I_1\left(|k|\sqrt{\frac{\sigma_z}{\sigma_\rho}} t\right) \tag{38}$$

$$B_{b2}(\rho, k) = i\frac{\mu_0 \sigma_z k}{\rho |k|}\sqrt{\frac{\sigma_\rho}{\sigma_z}}\left[\left[B_0(k)I_0\left(|k|\sqrt{\frac{\sigma_z}{\sigma_\rho}}t\right) + 2\sum_{n=1}^{\infty} B_n(k)I_n\left(|k|\sqrt{\frac{\sigma_z}{\sigma_\rho}}t\right)\right]\right.$$

$$\times \left[bI_1\left(|k|\sqrt{\frac{\sigma_z}{\sigma_\rho}}b\right) - tI_1\left(|k|\sqrt{\frac{\sigma_z}{\sigma_\rho}}t\right)\right]$$

$$+ \left[C_0 I_0\left(|k|\sqrt{\frac{\sigma_z}{\sigma_\rho}}t\right) + 2\sum_{n=1}^{\infty} C_n(k)I_n\left(|k|\sqrt{\frac{\sigma_z}{\sigma_\rho}}t\right)\right]$$

$$\left.\times \left[tK_1\left(|k|\sqrt{\frac{\sigma_z}{\sigma_\rho}}t\right) - bK_1\left(|k|\sqrt{\frac{\sigma_z}{\sigma_\rho}}b\right)\right]\right] \quad (39)$$

$$B_{b3}(\rho, k) = -i\frac{\mu_0 \sigma_z k}{\rho |k|}\sqrt{\frac{\sigma_\rho}{\sigma_z}}\left[B_0(k)aI_1\left(|k|\sqrt{\frac{\sigma_z}{\sigma_\rho}}a\right) + C_0(k)\left[\frac{1}{|k|}\sqrt{\frac{\sigma_\rho}{\sigma_z}} - aK_1\left(|k|\sqrt{\frac{\sigma_z}{\sigma_\rho}}a\right)\right]\right] \quad (40)$$

*Sheath:*

$$B_s(\rho, k) = i\frac{\mu_0 \sigma_s k}{\rho |k|}\left(D_0(k)(cI_1(|k|c) - bI_1(|k|)b) + E_0(k)(bK_1(|k|b) - cK_1(|k|)c)\right) \quad (41)$$

*Saline:*

$$B_e(\rho, k) = i\frac{\mu_0 \sigma_e k}{\rho |k|} F_0(k)(cK_1(|k|c) - \rho K_1(|k|)\rho) \quad (42)$$

Eventually, the amplitude of the MMG signal will be measured through highly sensitive magnetic sensors. The whole muscle can be assumed as a group of current-carrying line conductors along with the sensor sensitive axis. During the actual measurement, the Biot-Savart law can be utilized to perform minimum analysis calculations on these linear conductors to estimate the magnetic field strength where the magnetic flux density $\boldsymbol{B}(\boldsymbol{r})$ at the location $\boldsymbol{r}'$ can be expressed as below

$$\boldsymbol{B}(\boldsymbol{r}) = \frac{\mu_0 \mu_r I_{con}}{4\pi}\int \frac{d\boldsymbol{s}' \times (\boldsymbol{r} - \boldsymbol{r}')}{\|\boldsymbol{r} - \boldsymbol{r}'\|^3} \quad (43)$$

It is noted that the muscle fibre is locally regarded as a current-carrying conductor within a constant current, $I_{con}$, and an infinitely tiny length $d\mathbf{s}'$. The signal magnitude principally depends on the distance between a measurement point and the source, explained as $\mathbf{r} = r \cdot \mathbf{e}_R$. This cylindrical coordinate is suitable for further calculations with an assumption of the symmetrical conductor. With a linear conductor of length $l$ is along the z-axis with $\mathbf{r}' = z \cdot \mathbf{e}_z$, an infinitely short conductor cross-section, thus, can be defined as $d\mathbf{s}' = dz \cdot \mathbf{e}_z$. We finally obtain the magnetic flux density $\mathbf{B}(\mathbf{r})$:

$$\begin{aligned}
\mathbf{B}(\mathbf{r}) &= \mathbf{B}(r \cdot \mathbf{e}_R) \\
&= \frac{\mu_0 \mu_r I_{con}}{4\pi} \int_{z=0}^{l} \frac{dz \cdot \mathbf{e}_z \times (r \cdot \mathbf{e}_R - z \cdot \mathbf{e}_z)}{\|r \cdot \mathbf{e}_R - z \cdot \mathbf{e}_z\|^3} \\
&= \frac{\mu_0 \mu_r I_{con}}{4\pi} \int_{z=0}^{l} \frac{r \cdot dz \cdot \mathbf{e}_\varphi - z \cdot dz (\mathbf{e}_z \times \mathbf{e}_z)}{\sqrt{r^2 + z^2}^3} \\
&= \frac{\mu_0 \mu_r I_{con}}{4\pi} \int_{z=0}^{l} \frac{r \cdot dz \cdot \mathbf{e}_\varphi}{\sqrt{r^2 + z^2}^3} \\
&= \frac{\mu_0 \mu_r I_{con} r}{4\pi} \left[ \frac{z}{r^2 \sqrt{r^2 + z^2}} \right]_{z=0}^{l} \mathbf{e}_\varphi \\
&= \frac{\mu_0 \mu_r}{4\pi r} \cdot I_{con} \cdot \frac{l}{\sqrt{l^2 + r^2}} \mathbf{e}_\varphi
\end{aligned} \quad (44)$$

## Acknowledgements (optional)

Keep acknowledgements brief and do not include thanks to anonymous referees or editors, or effusive comments. Grant or contribution numbers may be acknowledged.

## Ethics declarations

Submission of a competing interests statement is required for all content of the journal.

## Supplementary Information

Supplementary Information: should be combined and supplied as a separate file, preferably in PDF format.